\documentclass[twocolumn,twocolappendix]{aastex63}

\received{}
\revised{}
\accepted{}

\submitjournal{The Astrophysical Journal}

\shorttitle{Gamma-ray \& Neutrino Emissions from Galaxy Clusters}
\shortauthors{Ha, Ryu, \& Kang}

\begin{document}

\title{Gamma-ray and Neutrino Emissions due to Cosmic-Ray Protons Accelerated at Intracluster Shocks in Galaxy Clusters}

\author[0000-0001-7670-4897]{Ji-Hoon Ha}
\affil{Department of Physics, School of Natural Sciences UNIST, Ulsan 44919, Korea}
\author[0000-0002-5455-2957]{Dongsu Ryu}
\affiliation{Department of Physics, School of Natural Sciences UNIST, Ulsan 44919, Korea}
\author[0000-0002-4674-5687]{Hyesung Kang}
\affiliation{Department of Earth Sciences, Pusan National University, Busan 46241, Korea}
\correspondingauthor{Dongsu Ryu}
\email{ryu@sirius.unist.ac.kr}

\begin{abstract}


We examine the cosmic-ray protons (CRp) accelerated at collisionless shocks in galaxy clusters using cosmological structure formation simulations. We find that in the intracluster medium (ICM) within the virial radius of simulated clusters, only $\sim7$\% of shock kinetic energy flux is dissipated by the shocks that are expected to accelerate CRp, that is, supercritical, quasi-parallel ($Q_\parallel$) shocks with sonic Mach number $M_s\ge2.25$. The rest is dissipated at subcritical shocks and quasi-perpendicular shocks, both of which may not accelerate CRp.  Adopting the diffusive shock acceleration (DSA) model recently presented in \citet{ryu2019}, we quantify the DSA of CRp in simulated clusters. The average fraction of the shock kinetic energy transferred to CRp via DSA is assessed at {$\sim(1-2)\times10^{-4}$}. {We also examine the energization of CRp through reacceleration using a model based on the test-particle solution. Assuming that the ICM plasma passes through shocks three times on average through the history of the universe and that CRp are reaccelerated only at supercritical $Q_\parallel$-shocks, the CRp spectrum flattens by $\sim0.05-0.1$ in slope and the total amount of CRp energy increases by $\sim40-80$\% from reacceleration.} We then estimate diffuse $\gamma$-ray and neutrino emissions, resulting from inelastic collisions between CRp and thermal protons. The predicted $\gamma$-ray emissions from simulated clusters lie mostly below the upper limits set by Fermi-LAT for observed clusters. The neutrino fluxes towards nearby clusters would be {$\lesssim10^{-4}$} of the IceCube flux at $E_{\nu}=1$ PeV and {$\lesssim10^{-6}$} of the atmospheric neutrino flux in the energy range of $E_{\nu}\leq1$ TeV.

\end{abstract}


\section{Introduction} 
\label{sec:s1}

During the formation of the large-scale structures (LSS) of the universe, shocks with low sonic Mach number of $M_s\lesssim 5$ are naturally induced by supersonic flow motions of baryonic matter in the hot intracluster medium \citep[ICM; e.g.,][]{miniati2000,ryu2003,pfrommer2006,skillman2008,vazza2009,schaal2015}. As in the cases of Earth's bow shock and supernova remnant shocks, these ICM shocks are collisionless, and hence are expected to accelerate cosmic-ray (CR) protons and electrons via diffusive shock acceleration \citep[DSA; e.g.,][]{bell1978,drury1983,kang2010,kang2013}. Giant radio relics such as the Sausage relic and the Toothbrush relic are interpreted as the structures of radio synchrotron emission from the CR electrons (CRe) accelerated at merger-driven ICM shocks \citep[see, e.g.,][and references therein]{vanweeren2019}. On the other hand, a clear confirmation of the acceleration of CR protons (CRp) in the ICM still remains elusive.

If CRp are produced at ICM shocks, owing to the long lifetime, most of them are expected be accumulated in galaxy clusters \citep[e.g.,][]{berezinsky1997}. Then, inelastic collisions between CRp with $E \gtrsim 1.22~{\rm GeV}$ (i.e., the threshold of the reaction) and thermal protons (CRp-p collisions) in the ICM produce neutral and charged pions, which decay through the following channels \citep[e.g.,][]{pfrommer2004}:
\begin{eqnarray}
\pi^0 &\rightarrow& 2\gamma, \nonumber\\ 
\label{pi0}
\pi^\pm &\rightarrow& \mu^\pm+\nu_{\mu}/\overline{\nu}_{\mu} \rightarrow e^\pm +\nu_{e}/\overline{\nu}_{e} +\nu_{\mu}+\overline{\nu}_{\mu}.
\label{piondecay}
\end{eqnarray}

The observation of diffuse cluster-wide $\gamma$-ray emission due to CRp-p collisions, hence, could provide an evidence for the production of CRp at ICM shocks. Such emission has been estimated with galaxy clusters from simulations for the LSS formation of the universe \citep[e.g.,][]{pinzke2010,zandanel2015,vazza2016}. However, currently available facilities such as Fermi-LAT so far have failed to detect $\gamma$-rays from clusters \citep[e.g.,][]{ackermann2014, ackermann2016}. Another evidence should be the detection of high-energy neutrinos, emitted by the same CRp-p collisions. For instance, \citet{murase2008,murase2013} estimated neutrinos due to the CRp produced at AGNs and SNRs in the ICM and cluster galaxies. \citet{zandanel2015} and \citet{murase2016}, on the other hand, suggested that ICM shocks and also accretion shocks surrounding clusters would not be the major sources of CRp that contribute significantly to the IceCube flux of neutrinos with $E_{\nu}\gtrsim10$ TeV.

Particle acceleration at collisionless shocks involves complex kinetic processes including micro-instabilities on various scales. It has been studied through particle-in-cell (PIC) and hybrid plasma simulations \citep[e.g.,][]{caprioli2014,guo2014,caprioli2015,park2015,ha2018b,kang2019}. The acceleration depends on several characteristics of collisionless shocks, such as the sonic ($M_s$) and Alfv\'en ($M_A$) Mach numbers, the plasma $\beta$ ($\equiv P_{\rm gas}/P_{\rm B}$, the ratio of gas to magnetic pressure), and the obliquity angle ($\theta_{\rm Bn}$), which is the angle between the shock normal and the mean magnetic field direction.

Collisionless shocks are classified as quasi-parallel ($Q_\parallel$) if $\theta_{\rm Bn}\lesssim 45^{\circ}$ and quasi-perpendicular ($Q_\perp$) if $\theta_{\rm Bn}\gtrsim 45^{\circ}$. CRp are known to be accelerated efficiently at $Q_\parallel$-shocks, while CRe are accelerated preferentially at $Q_\perp$-shocks \citep[e.g.,][]{marcowith2016}. Shocks associated with the solar wind have typically $\beta\sim1$ and $ 2\lesssim M_s\lesssim 10$ and supernova remnant shocks in the interstellar medium have $\beta\sim1$ and $M_s\lesssim 200$ \citep[e.g.,][]{kang2014}. On the other hand, ICM shocks are characterized with $\beta\sim 50-100$ and $M_s\lesssim 5$ \citep[e.g.,][]{ryu2003,ryu2008}. Although shocks with $\beta\sim 1$ have been extensively studied in the space-physics and astrophysics communities \cite[e.g.,][]{balogh13,marcowith2016}, {the accelerations of CRp and CRe at high-$\beta$ ICM shocks have been investigated only recently through PIC simulations \cite[e.g.,][]{guo2014, ha2018b, kang2019},} and has yet to be fully understood.

Since the efficacy of CRp production primarily governs CRp-p collisions, previous studies, where $\gamma$-ray emissions due to the CRp accelerated at shocks in galaxy clusters were estimated, adopted some recipes for the DSA efficiency \citep[e.g.,][]{pinzke2010,vazza2012,vazza2016}. The efficiency is often defined by the ratio of the postshock CRp energy flux, $F_{\rm CR}=E_{\rm CR}u_2$, to the shock kinetic energy flux, $F_{\phi}=E_{\rm sh}u_s=(1/2)\rho_1u_s^3$, as
\begin{equation}
\eta\equiv{F_{\rm CR}\over F_{\phi}}={1\over\chi}{E_{\rm CR}\over E_{\rm sh}}
\end{equation}
\citep{ryu2003}. Hereafter, the subscripts $1$ and $2$ denote the preshock and postshock states, respectively; $\rho$ and $u$ are the gas density and flow speed in the shock rest fame, $E_{\rm CR}$ is the postshock CRp energy density, $\chi=u_{\rm s}/u_2=\rho_2/\rho_1$ is the compression ratio across the shock, $E_{\rm sh}=(1/2)\rho_1 u_{\rm s}^2$ is the shock kinetic energy density, and $u_{\rm s}$ is the shock speed.

Based on fluid simulations of DSA where the time-dependent diffusion-convection equation for the isotropic part of CRp momentum distribution is solved along with a thermal leakage injection model, \citet{kang2013} suggested that $\eta$ could be as large as $\sim0.1$ for shocks with $M_s\simeq5$. According to the hybrid simulations performed by \citet{caprioli2014}, however, $\eta\approx0.036$ for the $M_s\approx 6.3$ ($M=5$ in their definition) shock in $\beta \sim 1$ plasmas. On the other hand, \cite{vazza2016} argued that the overall efficiency of CRp acceleration at ICM shocks with $2\lesssim M_s\lesssim 5$ should be limited to $\eta\lesssim10^{-3}$, if the predicted $\gamma$-ray emissions from simulated clusters are to be consistent with the upper limits set by Fermi-LAT for observed clusters \citep{ackermann2014}. This apparent discrepancy between the theoretical expectation and the observational constraint remains to be further investigated and is the main focus of this work.

Using PIC simulations, \citet{ha2018b} studied the injection and early acceleration of CRp at $Q_\parallel$-shocks with $M_s\approx2-4$ in hot ICM plasmas where $\beta\approx100$. They found that only supercritical $Q_\parallel$-shocks with $M_s\gtrsim2.25$ develop overshoot/undershoot oscillations in the shock transition, which lead to the specular reflection of incoming ions and further injection into the DSA process. Subcritical $Q_\parallel$-shocks with $M_s<2.25$, on the other hand, have relatively smooth structures, so the preaccleration and injection are negligible. Thus, $Q_\parallel$-shocks in the ICM may produce CRp only if $M_s\gtrsim2.25$.

Recently, \citet[][Paper I, hereafter]{ryu2019} proposed an analytic DSA model for supercritical $Q_\parallel$-shocks in the ICM that improves upon the test-particle DSA model for weak shocks described in \citet{kang2010}. The model incorporates the dynamical feedback of the CR pressure to the shock structure, and reflects the ``long-term'' evolution of the CRp spectrum in hybrid and PIC simulations \citep[e.g.,][]{caprioli2014,caprioli2015,ha2018b}. Based on the model, \citet{ryu2019} suggested that the DSA efficiency would be $\eta(M_s)\approx10^{-3}-10^{-2}$ for supercritical $Q_\parallel$-shocks with $M_s=2.25-5.0$.

{It was shown that the ICM gas passes through shocks more than once over the cosmological timescale \citep[see, e.g.,][]{ryu2003,vazza2009}. Hence, in addition to the production of CRp via DSA followed by ``fresh injection'', the previously produced CRp, which are transported along with the underlying ICM plasma throughout the cluster volume, could be further energized through ``reacceleration'' at subsequent shock passages. Although the reacceleration can substantially boost the CRp spectrum \citep[e.g.,][]{kang2011}, its importance in the ICM during the structure formation has not been evaluated quantitatively before.}

In this paper, by adopting the DSA model proposed in Paper I, we first estimate the CRp produced via fresh-injection DSA at ICM shocks in simulated sample clusters. {Assuming that those CRp fill the cluster volume and serve as the preexisting CRp, and adopting a simplified model for reacceleration based on the ``test-particle'' solution, we also estimate the boost of the CRp energy due to the multiple passages of the ICM plasma through shocks.} We then calculate $\gamma$-ray and neutrino emissions from simulated clusters using the approximate formalisms presented in \citet{pfrommer2004} and \citet{kelner2006}. The predicted $\gamma$-ray emissions are compared to the Fermi-LAT upper limits \citep{ackermann2014}. The neutrino fluxes from nearby clusters are compared with the IceCube flux \citep{aartsen2014} and the atmospheric neutrino flux \citep[e.g.,][]{richard2016}.

In Section \ref{sec:s2}, the estimation of CRp in simulated clusters is described. In Section \ref{sec:s3}, the calculation of $\gamma$-ray and neutrino emissions is presented. A brief summary follows in Section \ref{sec:s4}.

\section{CR protons in Simulated Clusters}
\label{sec:s2}

\subsection{Simulations and Galaxy Cluster Sample}
\label{sec:s2.1}

To generate a sample of simulated galaxy clusters, we performed a set of cosmological simulations, using a particle-mesh/Eulerian cosmological hydrodynamic code described in \citet{ryu1993}. The following parameters for a $\Lambda$CDM cosmology model were employed: baryon density $\Omega_{\rm BM}=0.044$, dark matter (DM) density $\Omega_{\rm DM}=0.236$, cosmological constant $\Omega_\Lambda=0.72$, Hubble parameter $h\equiv H_0/(100\ {\rm km}\ {\rm s}^{-1}{\rm Mpc}^{-1})=0.7$, rms density fluctuation $\sigma_8=0.82$, and primordial spectral index $n=0.96$. {These parameters are consistent with the WMAP7 data \citep[][]{komatsu2011}.} The simulation box has the comoving size of $57h^{-1}$ Mpc with periodic boundaries, and is divided into $1650^3$ grid zones, so the spatial resolution is $\Delta l = 34.5h^{-1}$ kpc. Nongravitational effects such as radiative and feedback processes were not considered; it was shown that the statistics of ICM shocks (see below) do not sensitively depend on nongravitational effects \citep[see, e.g.,][]{kang2007}. 

\begin{figure}[t]
\vskip 0.2 cm
\hskip 0 cm
\centerline{\includegraphics[width=0.52\textwidth]{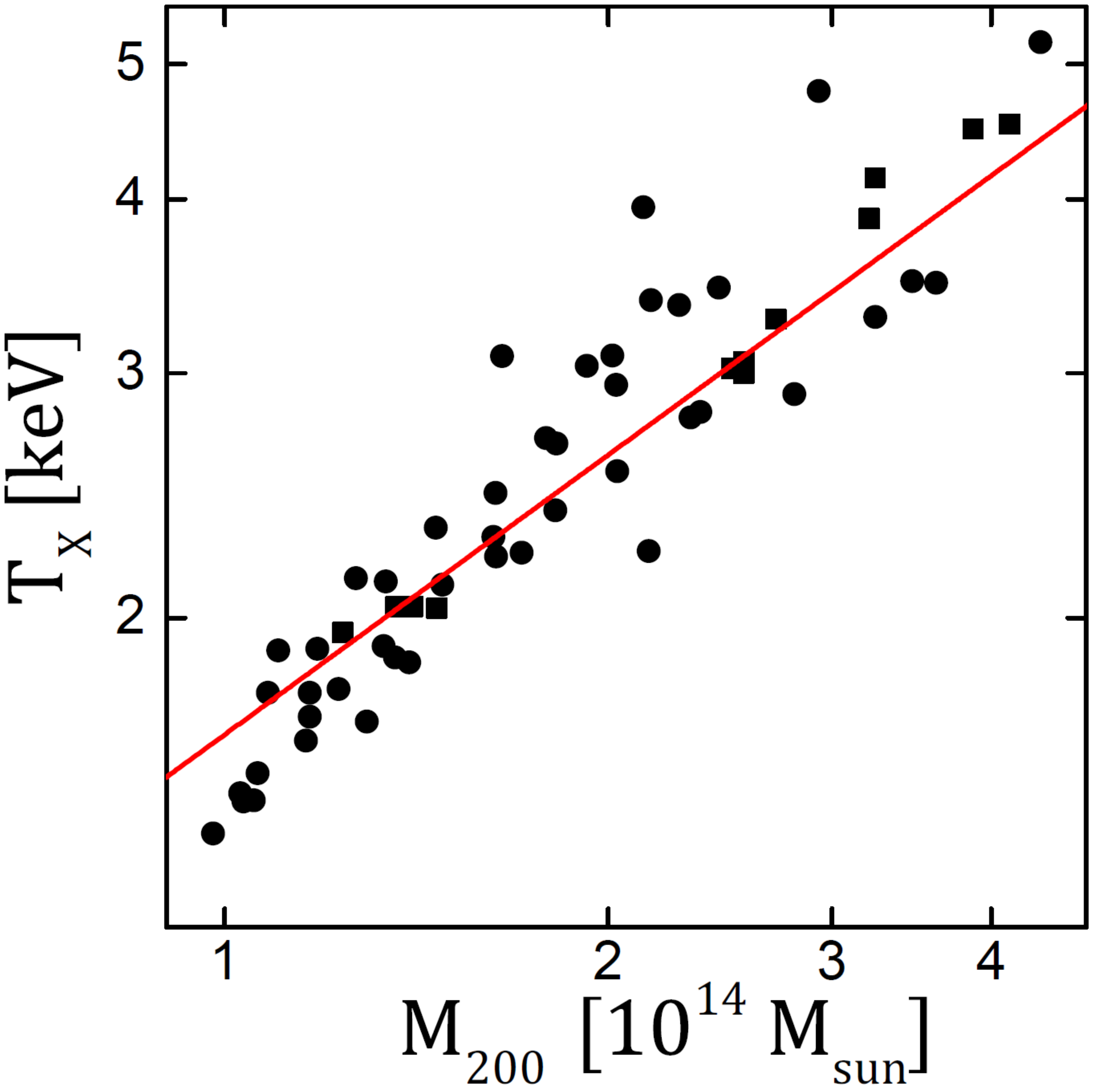}}
\vskip -0.2 cm
\caption{Mass vs temperature relation for 58 sample clusters at $z = 0$, found in four simulations for the LSS formation of the universe. The total (baryon plus DM) mass and the X-ray emission-weighted temperature inside the spherical volume of $r\le r_{200}$ are shown. The filled squares denote twelve clusters used to draw Figures \ref{fig:f2}. The red solid line represents the scaling relation of $T_{X}\propto M_{200}^{2/3}$.\label{fig:f1}}
\end{figure}

{The magnetic field, \mbox{\boldmath$B$}, which is necessary for differentiating between $Q_\parallel$ and $Q_\perp$-shocks (see Section \ref{sec:s2.2}), is assumed to be generated via the Biermann Battery mechanism at shocks \citep[][]{biermann1950}, and then advected passively. In our simulations, the following equation along with the equations for fluid and gravity were solved:
\begin{equation}
\frac{\partial\mbox{\boldmath$B$}}{\partial t} =\mbox{\boldmath$\nabla$}\times(\mbox{\boldmath$v$}\times\mbox{\boldmath$B$}) + \frac{c\nabla p_e\times\nabla n_e}{n_e^2 e},
\end{equation}
where $n_e$ and $p_e$ are the electron number density and pressure, respectively, and \mbox{\boldmath$v$} is the flow speed. The second term on the right hand side accounts for the Biermann battery mechanism. The passive evolution of \mbox{\boldmath$B$} implies that the Lorenz force term in the momentum equation is ignored, so the magnetic field does not affect the fluid motions. Further detailed descriptions can be found in \citet{kulsrude97}.}

In the simulation box, the local peaks of X-ray emissivity are identified as the centers of clusters, and the total (baryons plus DM) mass, $M_{200}$, and the X-ray emission-weighted temperature, $T_X$, of clusters inside $r_{200}$ are calculated \citep[e.g.,][]{kang1994}. Here, $r_{200}$ is the virial radius defined by the gas overdensity of $\rho_{\rm gas}/\langle\rho_{\rm gas}\rangle=200$. From the $z=0$ data of four simulations, a sample of 58 clusters with $1~{\rm keV} \lesssim T_X \lesssim 5~{\rm keV}$ are found. They have $10^{14} M_{\odot} \lesssim M_{200} \lesssim 5 \times 10^{14} M_{\odot}$ and $r_{200} \approx 1-2 h^{-1}{\rm Mpc}$. Figure \ref{fig:f1} shows the mass {\it versus} temperature relation of the sample clusters, which follows $T_{X} \propto M_{200}^{2/3}$, expected for virial equilibrium.

\subsection{Shock Identification}
\label{sec:s2.2} 

We identify ICM shocks formed inside simulated clusters, as follows \citep[see, e.g,,][]{ryu2003,hong2014}. Grid zones are defined as '`shocked'', if they meet the shock identification conditions along each principle axis: (1) $\mbox{\boldmath$\nabla$}\cdot\mbox{\boldmath$u$} < 0$, i.e., the converging local flow, (2) $\Delta T \times \Delta \rho > 0$, i.e., the same sign of the density and temperature gradients, and (3) $\left|{\Delta \log T}\right| > 0.11$, i.e., the temperature jump larger than that of $M_s=1.3$ shock. The shock transition typically spreads over $2 - 3$ zones in numerical simulations, and the zone with minimum $\nabla \cdot u$ is defined as the shock center. The sonic Mach number is calculated with the temperature jump across the shock transition as $T_2/T_1=(5M_s^{2}-1)(M_s^{2}+3)/(16M_s^{2})$. The Mach number of shock zones is defined as $M_s=$ max$(M_{s,x},M_{s,y},M_{s,z})$, where $M_{s,x}$, $M_{s,y}$, and $M_{s,z}$ are the Mach numbers along the principle axes. The shock speed is estimated as $u_s = M_s\sqrt{5P_{\rm gas,1}/3\rho_1}$. Shocks with $M_s\ge 1.5$ are identified, although only $Q_\parallel$-shocks with $M_s\ge2.25$ are accounted for the CRp production (see Sections \ref{sec:s2.3} and \ref{sec:s2.5}). Typically, a shock surface consists of a number of shock zones, and the surface area is estimated assuming each shock zone contributes $s_{\rm sh}=1.19 (\Delta l)^2$, which is the mean projected area of a zone for random shock normal orientation.

For shock zones, the shock obliquity angle is calculated as $\theta_{\rm Bn} \equiv \cos^{-1}[|\Delta\mbox{\boldmath$u$}\cdot\mbox{\boldmath$B$}_1|/(|\Delta\mbox{\boldmath$u$}||\mbox{\boldmath$B$}_1|)]$, where $\Delta\mbox{\boldmath$u$} = \mbox{\boldmath$u$}_2 - \mbox{\boldmath$u$}_1$ and $\mbox{\boldmath$B$}_1$ is the preshock magnetic field. Inside $r\leq r_{200}$ of simulated clusters, typically $\sim30\%$ of identified shock zones are $Q_\parallel$ with $\theta_{\rm Bn} \leq 45^{\circ}$, while the rest are $Q_\perp$ with $\theta_{B} > 45^{\circ}$ \citep[see also][]{wittor2017,roh2019}.

\subsection{CRp Production {via Fresh-Injection DSA}}
\label{sec:s2.3}

\begin{figure*}[t]
\vskip 0.2 cm
\hskip 0 cm
\centerline{\includegraphics[width=1.\textwidth]{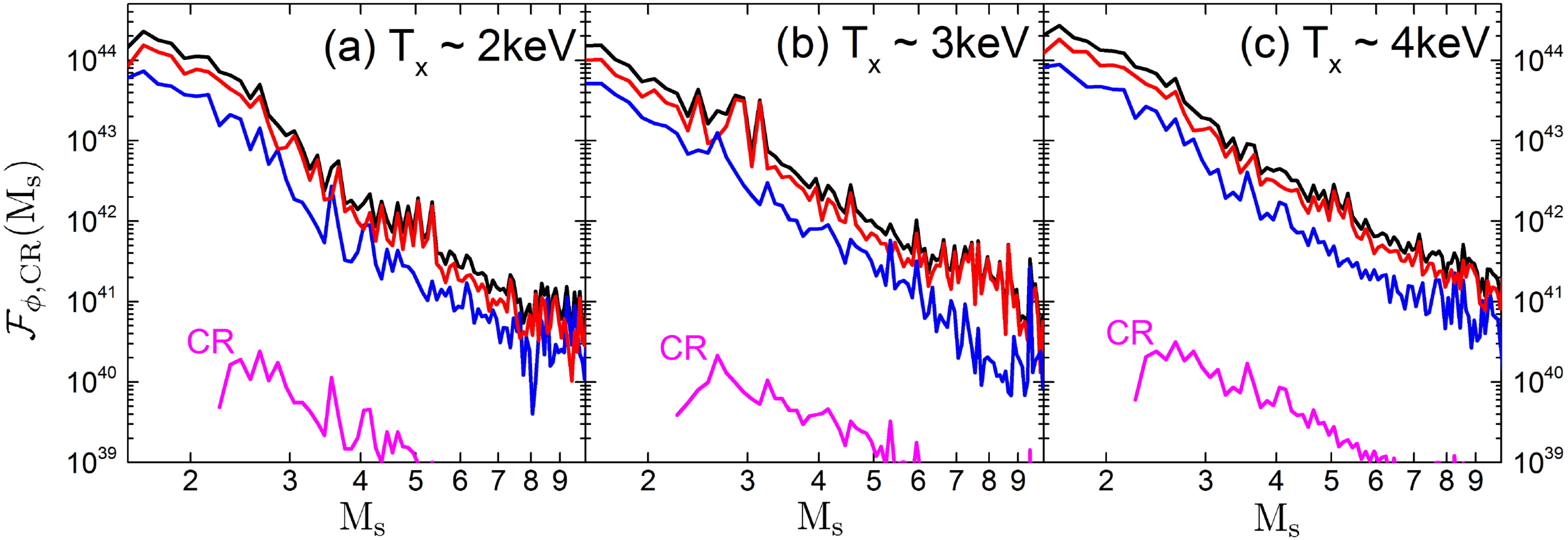}}
\vskip -0.1 cm
\caption{Shock kinetic energy flux, $\mathcal{F}_{\phi}$, and CRp energy flux, $\mathcal{F}_{\rm CR}$, in units of ${\rm erg\ s^{-1}}(h^{-1} {\rm Mpc})^{-3}$, as a function of $M_s$, processed through shocks inside the sphere of $r_{200}$ of sample clusters with the X-ray emission-weighted temperature (a) $T_X \sim 2$ keV, (b) $T_X \sim 3$ keV and (c) $T_X \sim 4$ keV. Each panel shows the fluxes averaged over four clusters with similar $T_X$, denoted with the filled squares in Figure \ref{fig:f1}. The black lines show $\mathcal{F}_{\phi}$ through all the shocks, while the red (blue) lines show $\mathcal{F}_{\phi}$ through $Q_\perp$ ($Q_\parallel$) shocks only. The magenta lines draw $\mathcal{F}_{\rm CR}$ produced by supercritical $Q_\parallel$-shocks.\label{fig:f2}}
\end{figure*}

To estimate the CRp produced via DSA, {followed by {\it in insu} injection at shock zones from the background thermal plasma}, we adopt the analytic model presented in Paper I. The main ideas of this model can be summarized as follows. (1) The proton injection and DSA are effective only at supercritical $Q_\parallel$-shocks with $M_s\gtrsim2.25$. (2) At weak $Q_\parallel$-shocks with $M_s\lesssim 5$, the postshock CR distribution, $f_{\rm CR}(p)$, follows the test-particle DSA power-law with the slope, $q=3\chi/(\chi-1)$, determined by the shock compression ratio, $\chi$. (3) The transition from the postshock Maxwellian to the CRp power-law distribution occurs at the so-called injection momentum, $ p_{\rm inj}$. The amplitude of $f_{\rm CR}(p)$ at $p_{\rm inj}$ is anchored at the thermal Maxwellian distribution. (4) As a fraction of the shock energy is transferred to CRp, the energy density of postshock thermal protons and hence the postshock temperature $T_2$ decrease self-consistently. At the same time, the normalization of $f_{\rm CR}(p)$ reduces. The weakening of the subshock due to the dynamical feedback of the CR pressure to the shock structure and the resulting reduction of $f_{\rm CR}(p)$ have been observed in numerical simulations \citep[e.g.,][Paper I]{kang2002,kang2005,caprioli2014}. (5) In the model, the CR energy density is kept to be less than 10 \% of the shock kinetic energy density for shocks with $M_s\lesssim5$, consistent with the test-particle treatment.

The analytic DSA model gives the momentum spectrum of CRp at shock zones as
\begin{equation}
f_{\rm CR}(p) \approx n_2 {{ ~\exp(-Q_{\rm i}^2)}\over \pi^{1.5}~p_{\rm th,p}^{3}} \left({p \over p_{\rm inj}}\right)^{-q}~~{\rm for}~p\geq p_{\rm inj},
\label{finj}
\end{equation}
for $Q_\parallel$-shocks with $M_s\ge2.25$. Here, $n_2$ and $p_{\rm th,p}\equiv\sqrt{2m_p k_B T_2}$ are the postshock number density and momentum of thermal protons, respectively, and $m_p$ is the proton mass, and $k_B$ is the Boltzmann constant. The injection momentum, $p_{\rm inj}$, is expressed in terms of the injection parameter, $Q_{\rm i}$, as
\begin{equation}
p_{\rm inj}= Q_{\rm i} \cdot p_{\rm th,p}.
\label{IP}
\end{equation}
In the model, $Q_{\rm i}=Q_{\rm i,0}/\sqrt{R_{\rm T}}$ with a fixed initial $Q_{\rm i,0}$ increases gradually, but approaches to an asymptotic value as the CR energy density increases. Considering the results from the hybrid simulations of \citet{caprioli2014} and \citet{caprioli2015} and the extended PIC simulation presented in Paper I, $Q_{\rm i,0}\approx 3.3-3.5$ is suggested. $R_{\rm T}$ is the reduction factor of the postshock temperature, which depends on both $M_s$ and $Q_{\rm i,0}$. Here, we present the production of CRp with $Q_{\rm i,0}=3.5$, along with $R_{\rm T}$ from Figure 4 of Paper I (see below for discussions on the dependence on $Q_{\rm i,0}$).

Then, the postshock energy density of CRp can be evaluated as
\begin{equation}
E_{\rm CR}= 4\pi c\int_{p_{\rm min}}^{\infty}(\sqrt{p^2+(m_{\rm p}c)^2}-m_{\rm p}c)f_{\rm CR}(p)\ p^2dp,
\label{ECR}
\end{equation}
where $c$ is the speed of light. For the lower bound of the integral, $p_{\rm min}=0.78\ {\rm GeV}/c$ is used, which is the threshold energy of $\pi$-production reaction. The postshock CRp energy flux is given as $F_{\rm CR}=E_{\rm CR}u_2$.

With the shock kinetic energy flux, $F_{\phi}=(1/2)\rho_1 u_s^3$, the DSA efficiency, $\eta(M_s)\equiv F_{\rm CR}(M_s)/F_{\phi}(M_s)$ (see the introduction), is given. The analytic DSA model of Paper I, adopted in this paper, suggests $\eta(M_s)\approx10^{-3}-10^{-2}$ for $Q_\parallel$-shocks with $M_s=2.25-5.0$. {Here, $F_{\rm CR}$ at ICM shocks is estimated using Equations (\ref{finj}) and (\ref{ECR}), rather than as $\eta(M_s)F_{\phi}$. However, for shocks with $M_s> 5$, which are beyond the Mach number range of the analytic DSA model (see Figure 4 of Paper I), $R_{\rm T}$ is adjusted, so that $F_{\rm CR}(M_s)/F_{\phi}(M_s)$ is limited to 0.01. We note that the contribution from shocks with $M_s> 5$ in the ICM is rather insignificant (see Figure \ref{fig:f2}).}

A few comments are in order. (1) In the case of weak shocks with low $M_s$, where the CRp spectrum is dominated by low-energy particles, the estimated $F_{\rm CR}$ depends rather sensitively on $p_{\rm min}$, although the $\pi$-production rate does not once $p_{\rm min}\leq0.78\ {\rm GeV}/c$. (2) If $Q_{\rm i,0}=3.3$, instead of $Q_{\rm i,0}=3.5$, is adopted, $F_{\rm CR}$ would be $\sim2$ times larger. (3) As mentioned in the introduction, \citet{kang2013} suggested $\eta(M_s)\sim0.1$ for $M_s\simeq5$, while \citet{caprioli2014} presented $\eta\approx0.036$ for $M_s\approx 6.3$. The analytic DSA model, adopted in this paper, assumes $\eta(M_s)$ and hence $F_{\rm CR}$, which are {about several to ten times smaller than those of \citet{kang2013} and \citet{caprioli2014}.}

\begin{figure*}[t]
\vskip 0.2 cm
\hskip -0.2 cm
\centerline{\includegraphics[width=0.75\textwidth]{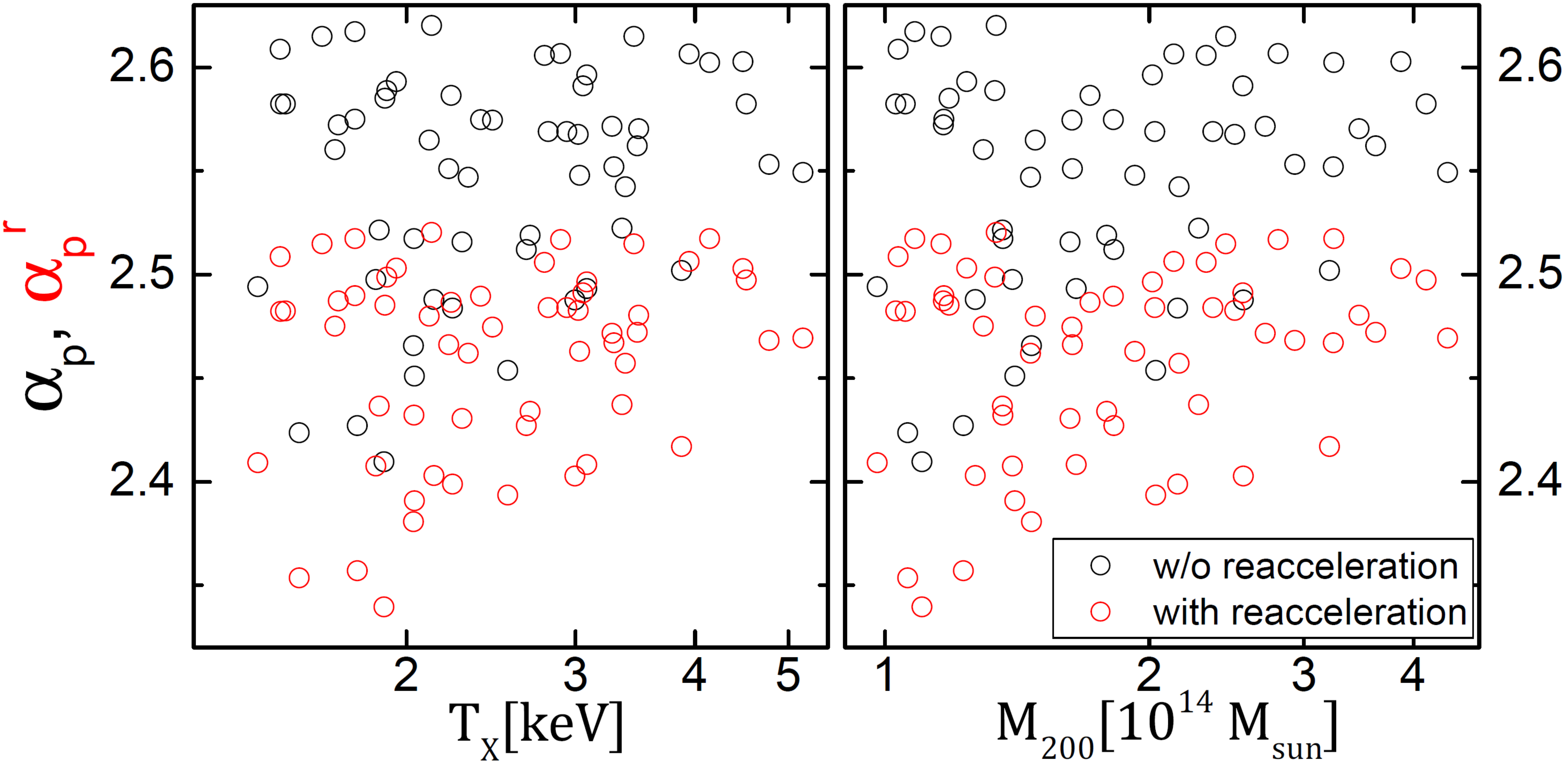}}
\vskip -0.1 cm
\caption{Slope of the volume-integrated CRp momentum distribution, produced by all supercritical $Q_{\parallel}$-shocks inside the sphere of $r_{200}$, as a function of the X-ray emission-weighted temperature and the total mass, for all 58 sample clusters. {The black and red open circles show the slopes without ($\alpha_p$) and with ($\alpha_p^r$) reacceleration incorporated, respectively.}\label{fig:f3}}
\end{figure*}

To quantify the CRp production at ICM shocks, we evaluate the energy flux processed through shocks inside sample clusters, as a function of the shock Mach number, as
\begin{equation}
\mathcal{F}_A(M_s)\ d\log M_s = \frac{1}{V_{<r_{200}}}\sum s_{\rm sh}F_A(M_s),
\end{equation}
where $A=\phi$ and $A={\rm CR}$ are used to denote the shock kinetic energy flux and the CRp energy flux, respectively. The summation goes over the shock zones with the Mach number between $\log M_s$ and $\log M_s+d\log M_s$ inside $r_{200}$, $V_{<r_{200}}=(4\pi/3)r_{200}^3$, and $s_{\rm sh}$ is the area of each shock zone. 
Figure \ref{fig:f2} shows $\mathcal{F}_{\phi}$ and $\mathcal{F}_{\rm CR}$ {at the present epoch ($z=0$)} for clusters with the X-ray emission-weighted temperature close to $T_X\sim2$ keV, $\sim3$ keV, and $\sim4$ keV. 
Weaker shocks dissipate a larger amount of shock kinetic energy, as pointed in previous works \citep[e.g.,][]{ryu2003,vazza2009}. Specifically, $\sim97\%$ of $\mathcal{F}_{\phi}$ is processed through shocks with $M_{s} \lesssim5$, and the fraction is not sensitive to cluster properties, such as $T_X$. We find that for all sample clusters, $\sim30\%$ of $\mathcal{F}_{\phi}$ is processed through $Q_\parallel$-shocks (blue lines) and the rest through $Q_\perp$-shocks (red lines); the partitioning is about the same as that of the frequency of $Q_\parallel$ and $Q_\perp$-shocks. Moreover, $\sim23\%$ of $\mathcal{F}_{\phi}$ associated with all $Q_\parallel$-shocks goes through supercritical shocks with $M_s\ge2.25$. As a result, only $\sim7$\%, or $\sim6-8$\% including the range for different clusters, of the shock kinetic energy is dissipated through supercritical $Q_{\parallel}$-shocks that are expected to accelerate CRp. 

Figure \ref{fig:f2} demonstrates that $\mathcal{F}_{\rm CR}$ (magenta lines), produced by supercritical $Q_\parallel$-shocks, is several orders of magnitude smaller than $\mathcal{F}_{\phi}$. We find that for all sample clusters, the total $\mathcal{F}_{\rm CR}$, integrated over $M_s$, is {$\sim(1-2)\times10^{-4}$} of the total $\mathcal{F}_{\phi}$. This can be understood as the average value of $\eta(M_s)\times\mathcal{F}_{\phi}(M_s)$, convoluted with the population of supercritical $Q_{\parallel}$-shocks. It means that the fraction of the shock kinetic energy transferred to CRp is estimated to be {$\sim(1-2)\times10^{-4}$}, based on the analytic DSA model adopted in this paper. If $Q_{\rm i,0}=3.3$ is used (the results are not shown), $\mathcal{F}_{\rm CR}$, and hence the amount of CRp produced, would be $\sim2$ times larger.

The number of CRp in the momentum bin between $p$ and $p+dp$, produced by ICM shocks, can be evaluated as follow:
\begin{equation}
{\dot\mathcal{N}}_{\rm CR}(p)\ dp = \sum\limits_{Q_{\parallel},\ M_s\ge2.25}4\pi s_{\rm sh}u_2f_{\rm CR}(p)\ p^2dp,
\label{NdotCR}
\end{equation}
where the summation includes the entire population of supercritical $Q_{\parallel}$-shocks with $M_s\ge2.25$ inside $r_{200}$. Note that ${\dot\mathcal{N}}_{\rm CR}(p)$ is defined in a way that $\int{\dot\mathcal{N}}_{\rm CR}(p)dp$ is the total rate of CRp production in the ICM. We fit it to a power-law, i.e., ${\dot\mathcal{N}}_{\rm CR}(p)\propto p^{-\alpha_p}$, with the volume-averaged slope $\alpha_p$. Figure \ref{fig:f3} show the values of $\alpha_p$, calculated for all 58 simulated galaxy clusters at $z=0$ (black open circles). The slope spreads over a range of $\alpha_p\sim2.4 - 2.6$, indicating that the average Mach number of the shocks of most efficient CRp production is in the range of $M_s\sim2.8-3.3$, {which is consistent with the Mach number range of large $\mathcal{F}_{\rm CR}$, $M_s\sim2.5-3.5$, in Figure \ref{fig:f2}}. We point that the slope in Figure \ref{fig:f3} is a bit larger than the values presented in \citet{hong2014} (see their Figure 10, where ${\bar q}=\alpha_p+2$). The difference can be understood with the difference in $\eta(M_s)$; $\eta(M_s=5)/\eta(M_s=2.25)$ is, for instance, $\sim10$ in the analytic model adopted in this paper, while it is $\sim100$ in the DSA efficiency model used in \citet{hong2014}. Hence, shocks with higher $M_s$ are counted with larger weights for the calculation of $\alpha_p$ in \citet{hong2014}.

\subsection{CRp Distribution in Sample Clusters}
\label{sec:s2.4}

Inside clusters, the CRp produced by ICM shocks are expected to be accumulated over the cosmological timescale, owing to their long lifetimes, as mentioned in the introduction. Although streaming and diffusion could be important for the transport of highest energy CRp, most of lower energy CRp should be advected along with the background plasma and magnetic fields \citep[e.g.,][]{ensslin2011,wiener2013,wiener2018}. Hence, the CRp distribution would be relaxed over the cluster volume via turbulent mixing on the typical dynamical timescale of the order of $\sim$ Gyr. Then, the total number of CRp in the momentum bin between $p$ and $p+dp$ accumulated inside clusters can be evaluated as
\begin{equation}
\mathcal{N}_{\rm CR}(p)=\int{\dot\mathcal{N}}_{\rm CR}(p)dt.
\label{Ncr}
\end{equation}

In our LSS formation simulations, we did not follow self-consistently in run-time the production of CRp at ICM shocks and their transport behind shocks. {Instead, we identify shocks and calculate $f_{\rm CR}(p)$ at shock zones in the post-processing step. We here attempt to approximate the above integral as
\begin{equation}
\mathcal{N}_{\rm CR}(p)\approx\tau_{\rm acc}\ {\dot\mathcal{N}}_{\rm CR}(p),
\label{Ncr2}
\end{equation}
with ${\dot\mathcal{N}}_{\rm CR}(p)$ estimated at $z=0$. Here, $\tau_{\rm acc}$ is the mean acceleration time scale. Note that the estimation of ${\dot\mathcal{N}}_{\rm CR}(p)$ at earlier epochs for a specific cluster found at $z=0$ is not feasible in post-processing, since the cluster has gone through a hierarchical formation history involving multiple mergers. Hence, in \citet{ryu2003}, \citet{skillman2008}, and \citet{vazza2009}, for instance, the shock population and the shock kinetic energy flux, $\mathcal{F}_{\phi}$, at different epochs were estimated, over the entire computational volume of LSS formation simulations, rather than inside the volume of a specific cluster. The Mach number distribution of $\mathcal{F}_{\phi}$ was presented in those studies; $\mathcal{F}_{\phi}(M)$ shows only a slow evolution from $z=1$ to 0, whereas it is somewhat smaller at higher redshifts.} By considering the time evolution of the shock population and the shock energy dissipation in LSS formation simulations, we use $\tau_{\rm acc}\sim5$ Gyr for all sample clusters. This approximation should give reasonable estimates within a factor of two or so.

Previous studies, in which the generation and transport of CRp were followed in run-time in LSS formation simulations, on the other hand, showed that CRp are produced preferentially in the cluster outskirts and then mixed, leading to the radial profile of the CR pressure, $P_{\rm CR}(r)$, which is broader than that of the gas pressure, $P_{\rm gas}(r)$ \citep[e.g.,][]{pfrommer2007,vazza2012,vazza2016}. This is partly because the shocks that can produce CRp ($M_s\gtrsim$ a few) are found mostly in the outskirts \citep[see, e.g.,][]{hong2014,ha2018a}, and also because the DSA efficiency is expected to increase with $M_s$ in the DSA theory \citep[see, e.g.,][Paper I]{kang2013}. Hence, we here employ an illustrative model for the radial profile of the CRp density that scales with the shell-averaged number density of gas particles as $n_{\rm CR}(r,p) \propto {\bar n_{\rm gas}(r)}^{\delta}$. We take $\delta = 0.5 - 1$, which covers most of the range suggested in the previous simulation studies cited above and observations \citep[e.g.,][]{brunetti2017}. Considering that the ICM is roughly isothermal, $\delta<1$ results in the radial profile of $P_{\rm CR}$ broader than that of $P_{\rm gas}$. For a smaller value of $\delta$, $n_{\rm CR}$ is less centrally concentrated, so the rate of inelastic CRp-p collisions occurring in the inner part of the cluster volume with high $n_{\rm gas}$ is lower.

{
\subsection{Energization of CRp through Reacceleration}
\label{sec:s2.5}

The ICM plasma passes through ICM shocks more than once, as mentioned in the introduction. The number of shock passages can be estimated with the amount of mass swept through shocks within the virial radius during $\tau_{\rm acc}$ as
\begin{equation}
N_{\rm passage} = \frac{\tau_{\rm acc}}{M_{\rm gas,200}}\sum s_{\rm sh}\rho_1u_s,
\end{equation}
where $M_{\rm gas,200}$ is the baryon mass inside $r_{200}$. Here, the summation goes over all the identified shock zones inside $r_{200}$. The value averaged for our 58 sample clusters is $\langle N_{\rm passage}\rangle\approx3.2$. Hence, the CRp produced via fresh-injection DSA during the first shock passage could be further energized by reacceleration, on average at two subsequent shock passages.

Here, we attempt to estimate the energization of CRp through reacceleration in the post-processing step, adopting the following ``simplified model''. It involves a number of assumptions, including the test-particle treatment for reacceleration, as follows.
(1) The ICM plasma passes through ICM shocks ``three times''. The three shock passages occur in sequence during each period of 
$\tau_{\rm acc}/3$, and hence the CRp production is a three-stage procedure. In the first stage, only fresh-injection DSA occurs. In the second and third stages, along with fresh-injection DSA, a fraction of the preexisting CRp, produced in the previous stages, is reaccelerated.
(2) Reacceleration operates only at supercritical $Q_\parallel$-shocks with $M_{\rm s} \ge 2.25$, as in the case of fresh-injection DSA. Even in the presence of preshock CRp, the reflection of protons at the shock front and the ensuing generation of upstream waves due to streaming protons is likely to be ineffective at subcritical shocks and $Q_\perp$-shocks. (See below for a discussion on the consequence of relaxing this assumption.)
(3)  With the preexisting CRp spectrum, $f_{\rm pre}(p)$, upstream of shock, the reaccelerated, downstream spectrum is given by the steady-state, test-particle solution as
\begin{equation}
f_{\rm reacc}(p) = q\ p^{-q}\int\limits_{p_{\rm inj}}^{p}p{\prime}^{q-1}f_{\rm pre}(p{\prime})dp{\prime},
\label{a1}
\end{equation}
where $q$ is again the test-particle power-law slope \citep[e.g.,][]{drury1983,kang2011}. In the case that $f_{\rm pre}(p)$ has a simple form, $f_{\rm reacc}(p)$ can be written down analytically (see Appendix \ref{sec:sa}).
(4) During each acceleration stage, CRp are advected and spread over $V_{<r_{200}}$, and the radial profile of the CRp density is described as $n_{\rm CR}(r) \propto {\bar n_{\rm gas}(r)}^{\delta}$ (see Section \ref{sec:s2.4}).

In the model, after the first stage, the CRp, produced solely via fresh-injection DSA and accumulated inside clusters, has the volume-integrated momentum distribution
\begin{equation}
\mathcal{N}^{\rm 1st}_{\rm CR}(p)\approx\frac{\tau_{\rm acc}}{3}\ {\dot\mathcal{N}}_{\rm CR}(p),
\label{Ncr-st1}
\end{equation}
where ${\dot\mathcal{N}}_{\rm CR}(p)$ is the CRp production rate in Equation (\ref{NdotCR}).

After the second stage, the volume-integrated CRp momentum distribution is given as
\begin{equation}
\mathcal{N}^{\rm 2nd}_{\rm CR}(p)\approx\left(\frac{2}{3}-\frac{\varphi}{3}\right){\tau_{\rm acc}}\ {\dot\mathcal{N}}_{\rm CR}(p)+\frac{\varphi}{3}{\tau_{\rm acc}}\ {\dot\mathcal{N}}_{\rm reacc}^{(1)}(p).
\label{Ncr-st2}
\end{equation}
Here, $\varphi$ is the fraction of preexisting CRp that passes through supercritical $Q_{\parallel}$-shocks and hence is reaccelerated. It may be inferred as
\begin{equation}
\varphi\approx\sum\limits_{Q_{\parallel},\ M_s\ge2.25}s_{\rm sh}\rho_1u_s {\Big/} \sum s_{\rm sh}\rho_1u_s,
\end{equation}
which is estimated to be $\varphi\sim6-8$\% for sample clusters. Note that $\varphi$ is almost identical to the fraction of the shock kinetic energy dissipated at supercritical $Q_{\parallel}$-shocks (see Section \ref{sec:s2.3}). 
${\dot\mathcal{N}}_{\rm reacc}^{(1)}(p)$ incorporates the reacceleration of CRp, and is estimated as follows. Assuming that the preexisting CRp produced in the first stage have a power-law momentum distribution, $f_{\rm pre}(p)\propto(p/p_{\rm inj})^{-s}$, and the radial density profile of $\propto{\bar n_{\rm gas}(r)}^{\delta}$, $f_{\rm reacc}^{(1)}(p)$ in Equation (\ref{a1}) is calculated at each supercritical $Q_{\parallel}$-shock zone; then all the contributions of reacceleration from shocks inside $r_{200}$ are added.

After the third, final stage, the volume-integrated CRp momentum distribution is given as
\begin{eqnarray}
\mathcal{N}^{\rm 3rd}_{\rm CR}(p)&&\approx\left(1-\varphi+\frac{\varphi^2}{3}\right){\tau_{\rm acc}}\ {\dot\mathcal{N}}_{\rm CR}(p) + \left(\varphi-\frac{2\varphi^2}{3}\right){\tau_{\rm acc}}\nonumber\\
&&\times{\dot\mathcal{N}}_{\rm reacc}^{(1)}(p)+\frac{\varphi^2}{3}{\tau_{\rm acc}}\ {\dot\mathcal{N}}_{\rm reacc}^{(2)}(p).
\label{Ncr-st3}
\end{eqnarray}
Here, ${\dot\mathcal{N}}_{\rm reacc}^{(2)}(p)$ represents the CRp that undergo the reacceleration twice. Similarly to ${\dot\mathcal{N}}_{\rm reacc}^{(1)}$, ${\dot\mathcal{N}}_{\rm reacc}^{(2)}$ is evaluated with $f_{\rm reacc}^{(2)}(p)$ in Equation (\ref{a2}).

\begin{figure}[t]
\vskip 0.2 cm
\hskip 0 cm
\centerline{\includegraphics[width=0.52\textwidth]{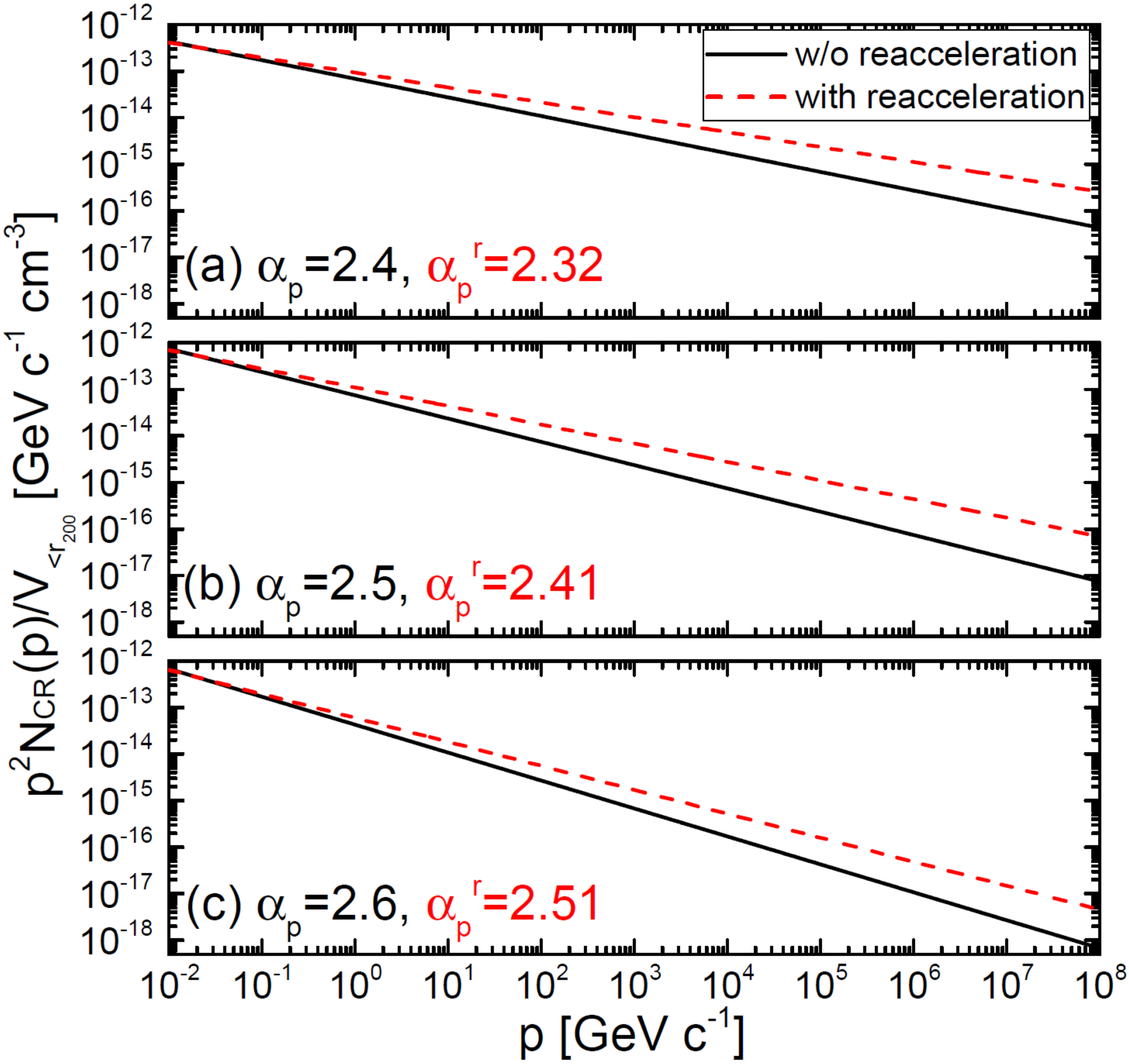}}
\vskip -0.2 cm
\caption{Volume-integrated CRp momentum distribution, $\mathcal{N}_{\rm CR}(p)$, produced by supercritical $Q_{\parallel}$-shocks inside the sphere of $r_{200}$, without (black solid lines) and with (red dashed lines) the energization of reacceleration, for three simulated clusters. Here, $\delta=0.75$ is used in the calculation of reacceleration. The volume-averaged slopes without and with reacceleration, $\alpha_p$ and $\alpha_p^r$, are given. \label{fig:f4}}
\end{figure}

\begin{figure*}[t]
\vskip 0.2 cm
\hskip -0.1 cm
\centerline{\includegraphics[width=1.\textwidth]{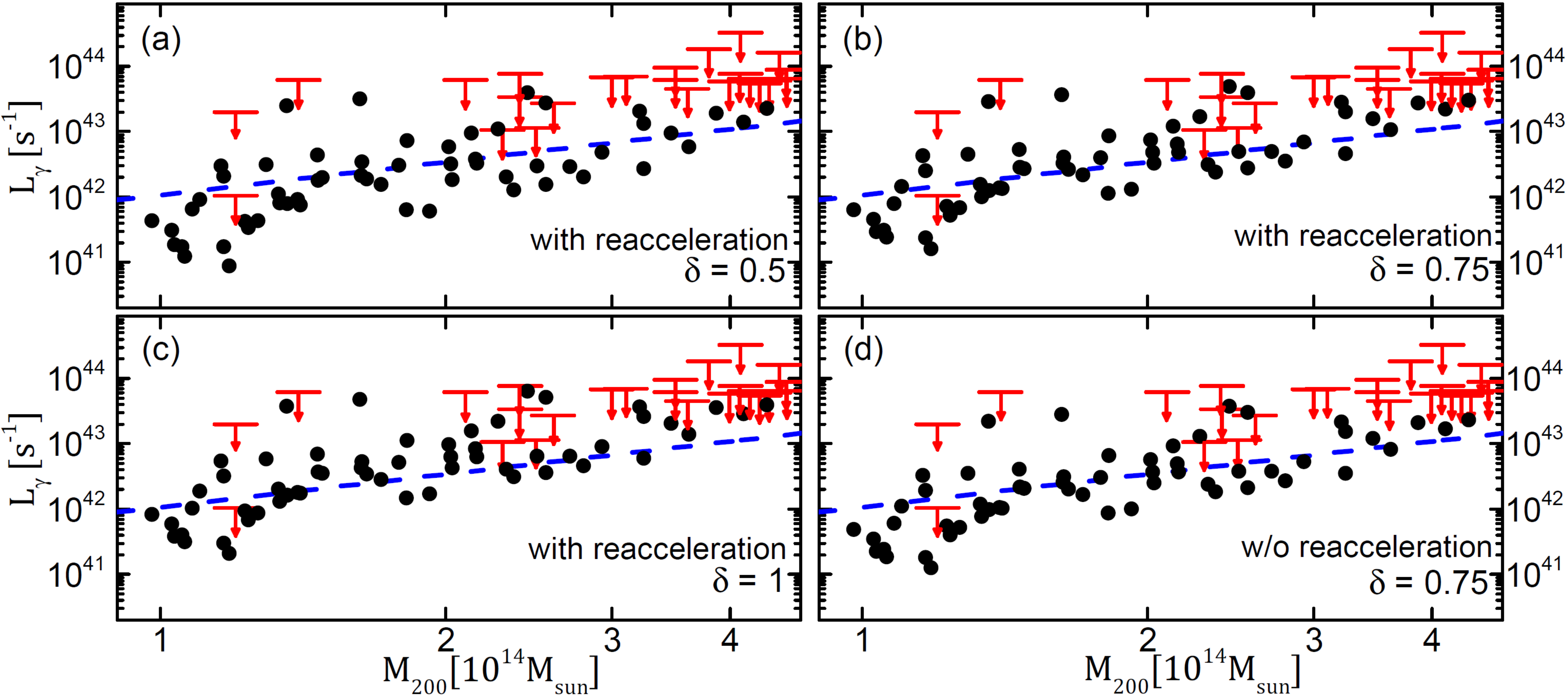}}
\vskip -0.1 cm
\caption{The number of $\gamma$-ray photons emitted per second in the energy band of [0.5, 200] GeV, $L_{\gamma}$, as a function of the total mass, for all 58 sample clusters (black circles). The red horizontal bars are the upper limits for observed clusters by Fermi LAT. The blue dashed lines draw the mass-luminosity relation, $L_{\gamma}\propto M_{200}^{5/3}$, assuming virial equilibrium and a constant CRp-to-gas energy ratio. {The panels (a) - (c) show $L_{\gamma}$ estimated from the CRp with reacceleration incorporated;} the three panels are for the different spatial distribution models of CRp with different $\delta$. {The panel (d) shows $L_{\gamma}$ from the CRp without reacceleration for $\delta=0.75$, for comparison.}\label{fig:f5}}
\end{figure*}

In Figure \ref{fig:f4}, the volume-integrated momentum distributions without (Equation (\ref{Ncr2})) and with (Equation (\ref{Ncr-st3})) the energization of reacceleration are compared for three simulated clusters at $z=0$; $\delta=0.75$ is used in the calculation of reacceleration contribution. Reacceleration conserves the number of CRp, and hence, the total number of CRp, $\int\mathcal{N}_{\rm CR}(p)dp$, remains the same. On the other hand, it makes the momentum spectrum harder, that is, $\mathcal{N}_{\rm CR}(p)$ becomes flatter, as shown in Figure \ref{fig:f4}. For all sample clusters, $\mathcal{N}_{\rm CR}(p)$ in Equation (\ref{Ncr-st3}) including the energization of reacceleration is again fitted to a power-law form with the slope, $\alpha_p^r$. In Figure \ref{fig:f3}, the estimated values of $\alpha_p^r$ are compared to those without reacceleration, $\alpha_p$; $\alpha_p^r \sim 2.35 - 2.5$, while $\alpha_p\sim2.4-2.6$ (see Section \ref{sec:s2.3}), that is, the momentum spectrum flattens by $\sim0.05-0.1$ due to reacceleration.

A flatter spectrum means a larger number of high energy CRp, and hence, the total energy contained in the CRp component (see Equation (\ref{ECR})) should be larger. We find that the total CRp energy increases due to reacceleration by $\sim40-80$\% with a mean value of $\sim60$\% when averaged for all clusters, if $\delta=0.75$ is assumed; the averaged increment is $\sim75$\% and $\sim50$\% for $\delta=0.5$ and 1, respectively. This number can be understood as follows. In Equation (\ref{Ncr-st3}), the major contribution of reacceleration is included in the $\varphi{\dot\mathcal{N}}_{\rm reacc}^{(1)}(p)$ term. The boost of the CRp energy with Equation (\ref{a1}) is, for instance, $\lesssim10$ for shocks with $M_s\sim3$ \citep[see Figure 2 of][]{kang2011}, while $\varphi\sim6-8$\%. 

For completeness, a few additional numbers are given here. If reacceleration operates at all (both supercritical and subcritical) $Q_{\parallel}$-shocks, the total CRp energy contained in sample clusters increases by $\sim90$\% on average (when $\delta=0.75$ is assumed). If reacceleration were to operate at all shocks, that is, both $Q_{\parallel}$ and $Q_{\perp}$-shocks, then the CRp energy would be increased by several times, which is probably too large to be compatible with the Fermi upper limits (see Section \ref{sec:s3}).
}

Below, for the estimations of $\gamma$-ray and neutrino emissions, we use the CRp expressed as
\begin{equation}
n_{\rm CR}(r,p)\ dp \approx n_{\rm CR0}\left[\frac{{\bar n_{\rm gas}}(r)}{n_{\rm gas}(0)}\right]^{\delta}\left(\frac{p}{{\rm GeV}/c}\right)^{-\alpha_p^r}\frac{dp}{{\rm GeV}/c},
\label{nrp}
\end{equation}
where $n_{\rm gas}(0)$ is the gas particle number density at the cluster center. The normalization factor, $n_{\rm CR0}$, is fixed by the condition
\begin{equation}
\int\limits_{<r_{200}}\int n_{\rm CR}(r,p)\ dp\ dV = \int\mathcal{N}_{\rm CR}(p)\ dp,
\label{norm}
\end{equation}
where the volume integral is over the sphere inside $r_{200}$.

\section{Gamma-Rays and Neutrinos from Simulated Clusters}
\label{sec:s3}

{In this section, we calculate $\gamma$-ray and neutrino emissions from simulated clusters, using $n_{\rm CR}(r,p)$ in Equation (\ref{nrp}), which includes the energization due to reacceleration. 
To speculate the consequence of reacceleration, we first compare the numbers of CRp with and without reacceleration, in the two momentum ranges: (1) $p_{\rm min}<p<10^3\ {\rm GeV}/c$ ($p_{\rm min}=0.78\ {\rm GeV}/c$) where most of the $\gamma$-rays observed by Fermi-LAT in the energy band of [0.5, 200] GeV are produced, and (2) $10^6<p<10^8\ {\rm GeV}/c$ where most of the high-energy neutrinos detected by IceCube are produced (see below). 
The number of CRp in $p_{\rm min}<p<10^3\ {\rm GeV}/c$ is increased by $\sim1.3-2.5$ times, while that in $10^6<p<10^8\ {\rm GeV}/c$ is increased by $\sim4.1-6.8$ times, due to reacceleration. Hence, reacceleration would have a limited consequence on the $\gamma$-rays observation with Fermi-LAT. On the other hand, it substantially boosts high-energy neutrinos from clusters.}

\subsection{Gamma-Ray Emissions}
\label{sec:s3.1}

For the calculation of $\gamma$-ray emissions from simulated clusters, we employ the approximate formula for the $\gamma$-ray source function as a function of $\gamma$-ray energy $E_{\gamma}$, presented in \citet{pfrommer2004};
\begin{eqnarray}
&&q_{\gamma}(r, E_{\gamma})dE_{\gamma}dV \approx c\sigma_{\rm pp}{\bar n_{\rm gas}}(r)\tilde{n}_{\rm CR}(r)\frac{2^{4-\alpha_{\gamma}}}{3\alpha_{\gamma}}\nonumber\\
&&\times\left(\frac{m_{\pi^0}c^2}{\rm GeV}\right)^{-\alpha_{\gamma}}\left[\left(\frac{2E_{\gamma}}{m_{\pi^0}c^2}\right)^{\delta_{\gamma}}+\left(\frac{2E_{\gamma}}{m_{\pi^0}c^2}\right)^{-\delta_{\gamma}}\right]^{-\frac{\alpha_{\gamma}}{\delta_{\gamma}}}\nonumber\\
&&\times\frac{dE_{\gamma}}{\rm GeV}dV,
\end{eqnarray}
where $\alpha_{\gamma}=4/3(\alpha_p^r-1/2)$ is the slope of $\gamma$-ray spectrum, $\delta_{\gamma}=0.14\alpha_{\gamma}^{-1.6} + 0.44$ is the shape parameter, $\sigma_{\rm pp}=32\times(0.96+e^{4.4-2.4\alpha_{\gamma}})$ mbarn is the effective cross-section of inelastic CRp-p collision, and $m_{\pi^0}$ is the pion mass. In our model, $\tilde{n}_{\rm CR}(r)=n_{\rm CR0}[\bar n_{\rm gas}(r)/n_{\rm gas}(0)]^{\delta}$. Then, the number of $\gamma$-ray photons emitted per second from a cluster is given as 
\begin{equation}
L_{\gamma}=\int_{<r_{200}} \int_{E1}^{E2} q_{\gamma}(r, E_{\gamma})\ dE_{\gamma}dV.
\end{equation}

Using ${\bar n_{\rm gas}}(r)$ and $\alpha_p^r$ calculated for simulated clusters with $\delta = 0.5$, 0.75, and 1, we estimate $L_{\gamma}$ of 58 sample clusters. The energy band of [$E1$, $E2$] = [0.5, 200] GeV is used to compare the estimates with the Fermi-LAT upper limits presented in \citet{ackermann2014}. Figure \ref{fig:f5} shows the estimates for $L_{\gamma}$ as a function of the cluster mass $M_{200}$, along with the Fermi-LAT upper limits. A few points are noted. (1) Because clusters with similar masses may undergo different dynamical evolutions, they could experience different shock formation histories and have different CRp productions. Hence, the $L_{\gamma}-M_{200}$ relation exhibits significant scatters. (2) Assuming virial equilibrium and a constant CRp-to-gas energy ratio, the mass-luminosity scaling relation, $L_{\gamma} \propto M_{200}^{5/3}$, is predicted \citep[see, e.g.,][]{pinzke2010, zandanel2015,vazza2016}. Although there are substantial scatters, $L_{\gamma}$'s for our sample clusters seem to roughly follow the predicted scaling relation. (3) Different CRp spatial distributions with different $\delta$ give different estimates for $L_{\gamma}$ within a factor of two (see the panels (a), (b), and (c)). Being the most centrally concentrated, the model with $\delta=1$ produces the largest amount of $\gamma$-ray emissions. 
{(4) The panels (b) and (d) compare $L_{\gamma}$'s from the CRp with and without reacceleration boost ($\delta=0.75$). As speculated above, the difference in $L_{\gamma}$'s is small, indicating that estimated $L_{\gamma}$ is not sensitive to whether the reacceleration of CRp at ICM shocks is included or not.}

All the models shown in Figure \ref{fig:f5}, including the one with reacceleration for $\delta=1$, result in $L_{\gamma}$'s that are mostly below the Fermi-LAT upper limits. Hence, although there are uncertainties in our estimation for the production of CRp at ICM shocks, we conclude that the DSA model proposed in Paper I is consistent with the Fermi-LAT upper limits.  

We attempt to compare our results with the predictions made by \citet{vazza2016}, in particular, the one for their CS14 model of the DSA efficiency, $\eta_{\rm CS14}(M_s)$, which adopted the efficiency based on the hybrid simulations of \citet{caprioli2014} for high $M_s$ along with the fitting form of \citet{kang2013} for the $M_s$ dependence in low $M_s$. For instance, the red triangles (labeled as CS14) in Figure 7 of \citet{vazza2016} shows $L_{\gamma}\approx 2-4\times 10^{43}\ {\rm photons\ s}^{-1}$ for simulated clusters with $M_{200} \approx 2-3\times 10^{14} M_{\odot}$, while our estimates for the model with $\delta=0.75$ vary as $L_{\gamma}\approx 0.5-2\times 10^{43}\ {\rm photons\ s}^{-1}$ for the same mass range. The ICM shock population and energy dissipation should be similar in the two works \citep[see, e.g.,][]{ryu2003,vazza2009}; {also the fraction of $Q_\parallel$-shocks is $\sim 30$\% in both works \citep[see][and Section \ref{sec:s2.2}]{wittor2017}.} One of differences in the two modelings is that for subcritical $Q_{\parallel}$-shocks with $M_s<2.25$, we assume no production of CRp at all, while $\eta_{\rm CS14}(M_s)$ is not zero. However, this may not lead to a significant difference in the CRp production, since $\eta_{\rm CS14}(M_s)$ sharply decreases with decreasing $M_s$ in the regime of $M_s\lesssim3$. On the other hand, with the DSA model adopted here, $\eta(M_s)\approx10^{-3}-10^{-2}$ for $M_s=2.25-5$ (see Section \ref{sec:s2.3}), which is lower by up to a factor of three to four times than $\eta_{\rm CS14}(M_s)$, explaining the difference in the predicted $L_{\gamma}$ in the two studies.

\begin{figure}[t]
\vskip 0.2 cm
\hskip 0 cm
\centerline{\includegraphics[width=0.52\textwidth]{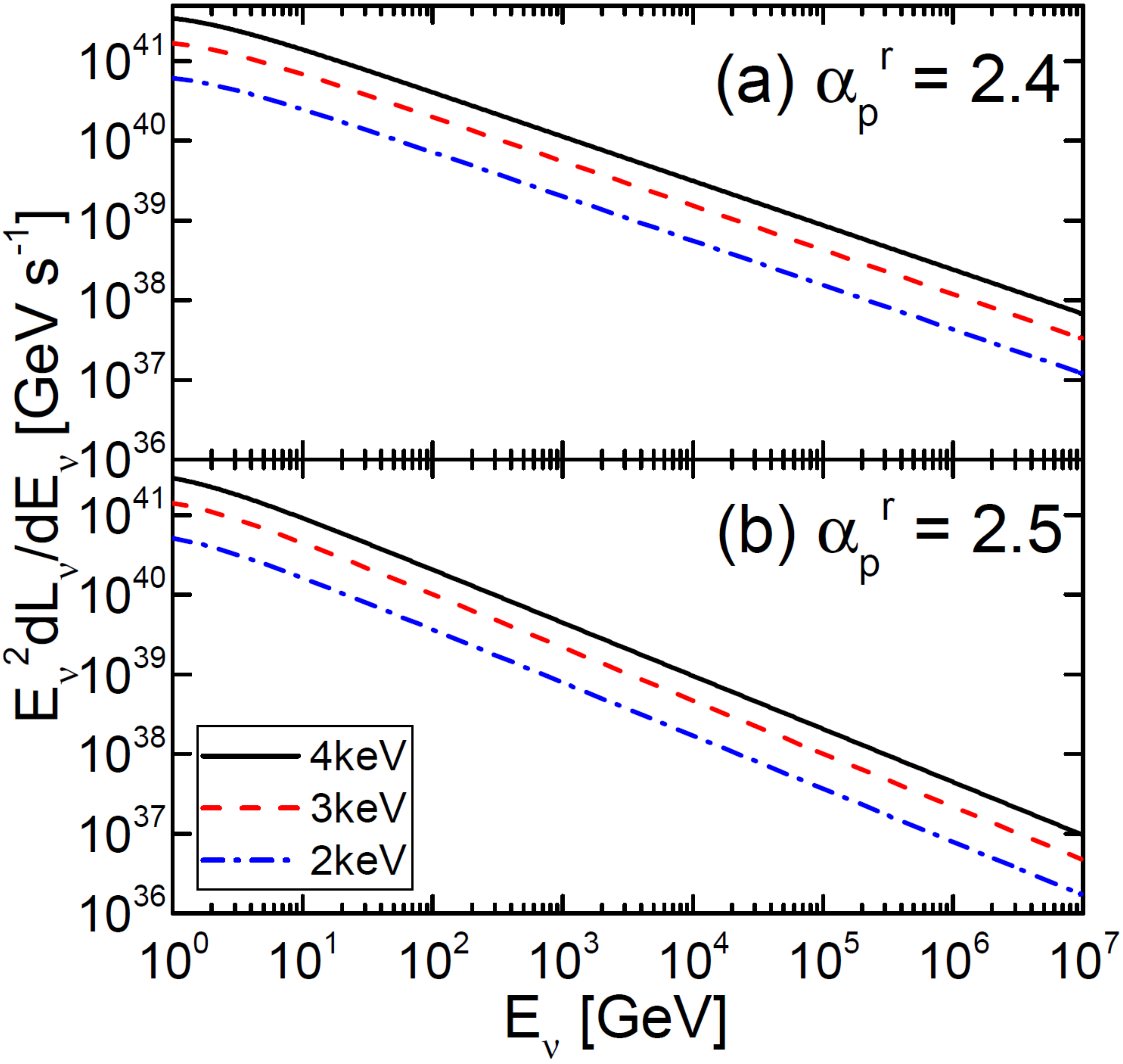}}
\vskip -0.2 cm
\caption{Energy spectrum of neutrinos from the sample clusters of $T_X\sim 2$ keV (blue dashed dot lines), 3 keV (red dashed lines), and 4 keV (black solid lines). Each line shows the spectrum averaged over four clusters with similar $T_X$. For the CRp distribution, {$\alpha_p^r=2.4$} and $\delta=0.75$ are used in the upper panel, and {$\alpha_p^r=2.5$} and $\delta=0.75$ in the lower panel, respectively.\label{fig:f6}}
\end{figure}

\subsection{Neutrino Emissions}
\label{sec:s3.2}

To calculate neutrino emissions from simulated clusters, we employ the analytic prescription described in \citet{kelner2006}. Assuming that the pion source function as a function of pion energy $E_{\pi}$ has a power-law form, $q_{\pi}(r,E_{\pi}) \propto E_{\pi}^{-\alpha_{\gamma}}$, the neutrino source function at the neutrino energy $E_{\nu} = E_{\gamma}$ is approximately related to the $\gamma$-ray source function as
\begin{equation}
q_{\nu}(r,E_{\nu}) = q_{\gamma}(r,E_{\gamma})[Z_{\nu_{\mu}}(\alpha_{\gamma})+Z_{\nu_{e}}(\alpha_{\gamma})].
\end{equation}
Here, 
\begin{eqnarray}
Z_{\nu_{\mu}}(\alpha_{\gamma}) &=& \frac{4[3-2k-k^{\alpha_{\gamma}}(3-2k+\alpha_{\gamma}-k\alpha_{\gamma})]}{\alpha_{\gamma}(1-k)^2(\alpha_{\gamma}+2)(\alpha_{\gamma}+3)} \nonumber\\
&+& (1-k)^{\alpha_{\gamma}-1}, \\
Z_{\nu_{e}}(\alpha_{\gamma}) &=& \frac{24[(1-k)\alpha_{\gamma}-k(1-k^{\alpha_{\gamma}})]}{\alpha_{\gamma}(1-k)^2(\alpha_{\gamma}+1)(\alpha_{\gamma}+2)(\alpha_{\gamma}+3)},
\end{eqnarray}
with $k=m_{\mu^{\pm}}^2/m_{\pi^{\pm}}^2=0.573$ account for the contributions of muon and electron neutrinos, respectively. Then, the energy spectrum of neutrons emitted per second from a cluster is estimated by
\begin{equation}
\frac{dL_{\nu}}{dE_{\nu}} = \int_{<r_{200}}q_{\nu}(r,E_{\nu})\ dV.
\end{equation}

Figure \ref{fig:f6} plots $E_{\nu}^2dL_{\nu}/dE_{\nu}$ as a function of $E_{\nu}$ for simulated clusters; the lines with different colors are for the sample clusters with $T_X$ close to $\sim2$ keV, $\sim3$ keV, and $\sim4$ keV, respectively. The upper and lower panels show the estimated spectra for the volume-averaged slope of CRp momentum distribution, {$\alpha_p^r=2.4$ and 2.5, respectively, which cover the range of $\alpha_p^r$ of simulated clusters (see Figure \ref{fig:f3})}; for the spatial distribution of CRp, $\delta=0.75$ is used. The spectrum has the energy dependence of {$\propto E_{\nu}^{-2.53}$ for $\alpha_p^r\sim 2.4$ and $\propto E_{\nu}^{-2.67}$ for $\alpha_p^r\sim 2.5$}, according to $\alpha_{\gamma}=4/3(\alpha_p^r-1/2)$. The number of neutrinos emitted from clusters of $T_X\sim2-4$ keV is estimated to be {$\sim10^{33}-10^{34}\ {\rm GeV}^{-1}{\rm s}^{-1}$ at $E_{\nu}\sim1$ TeV and $\sim$ a few $\times(10^{24}-10^{26})\ {\rm GeV}^{-1}{\rm s}^{-1}$ at $E_{\nu}\sim1$ PeV.}

\begin{deluxetable}{ccccccc}[t]
\tablecaption{List of Nearby Clusters \label{tab:t1}}
\tabletypesize{\small}
\tablenum{1}
\tablehead{
\colhead{Cluster} & \colhead{~} &
\colhead{$d$ [Mpc] $^a$} & \colhead{~} &
\colhead{$T_X$ [keV] $^b$}& \colhead{~} &
\colhead{$R_{\rm vir}$ [Mpc] $^b$}
}
\startdata
Virgo     & & 16.5  & & 2.3   & & 1.08  \\
Centaurus & & 41.3  & & 3.69  & & 1.32  \\
Perseus   & & 77.7  & & 6.42  & & 1.58  \\
Coma      & & 102   & & 8.07  & & 1.86 \\
Ophiuchus & & 121   & & 10.25 & & 2.91 \\
\enddata
\tablenotetext{a}{References for the cluster distances: \citet{mei2007} for the Virgo cluster, \citet{mieske2003} for the Centaurus cluster, \citet{Aleksic2012} for the Perseus cluster, \citet{thomsen1997} for the Coma cluster, and \citet{durret2015} for the Ophiuchus cluster.}
\tablenotetext{b}{The X-ray temperature and virial radius of the Virgo cluster are from \citet{urban2011}. Those of the Centaurus, Perseus, Coma, and Ophiuchus clusters are from \citet{chen2007}.}
\vspace{-0.8cm}
\end{deluxetable}

\begin{figure}[t]
\vskip 0.2 cm
\hskip 0 cm
\centerline{\includegraphics[width=0.52\textwidth]{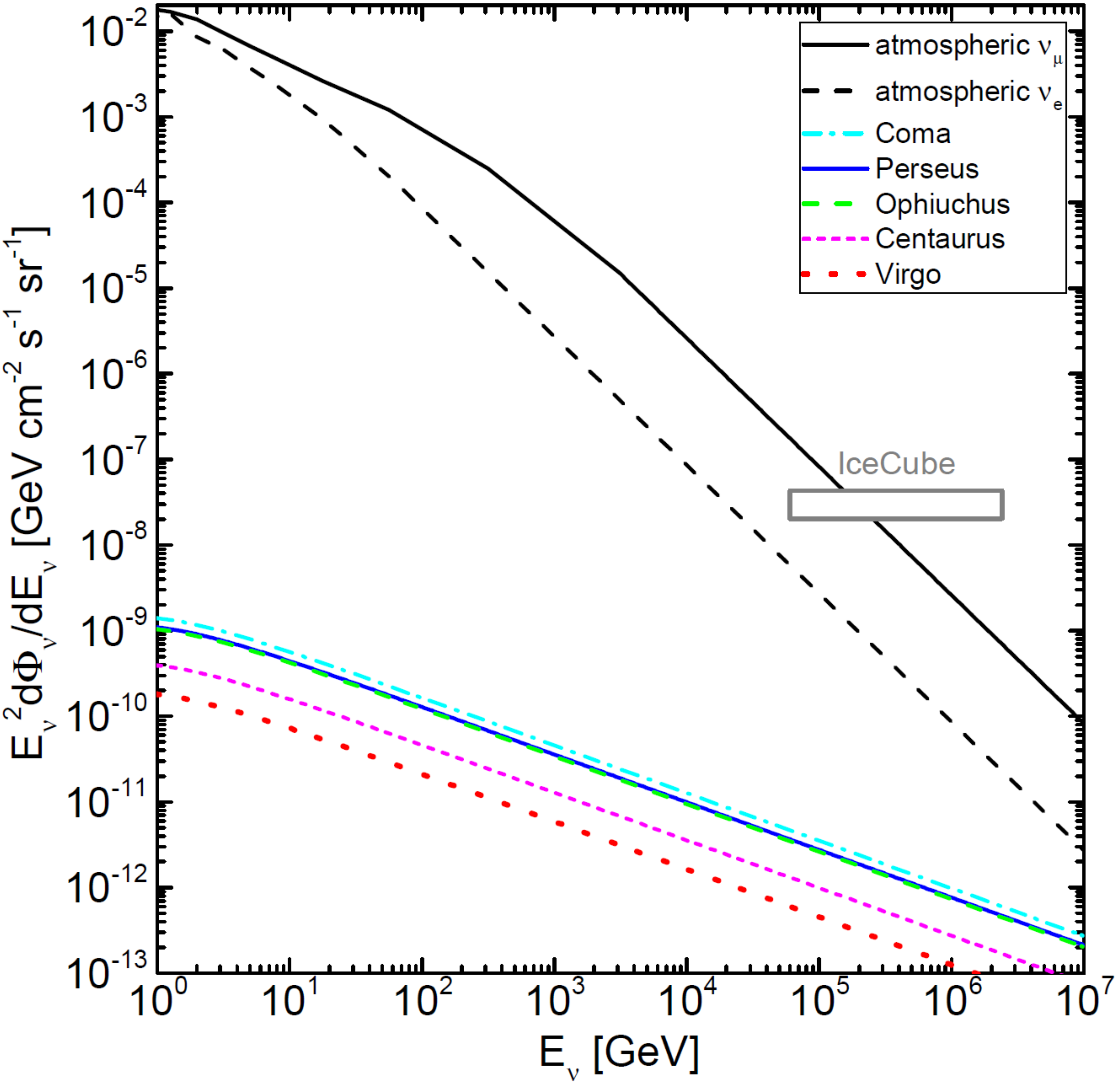}}
\vskip -0.2 cm
\caption{Predicted neutrino fluxes from nearby clusters. For the CRp distribution, the model with {$\alpha_p^r=2.4$} and $\delta=0.75$ is used. The gray box denotes the IceCube flux \citep[][]{aartsen2014}, and the black solid and dashed lines draw the fluxes of atmospheric muon and electron neutrinos \citep[][]{richard2016}.\label{fig:f7}}
\end{figure}

We also try to assess neutrino fluxes from the five nearby clusters listed in Table \ref{tab:t1}. Due to the limited box size of the LSS formation simulations here, the parameters of our sample clusters (see Figure \ref{fig:f1}) do not cover those of some of the nearby clusters. Hence, we employ the scaling relation $L_{\nu}\propto T_X^{5/2}$, along with the neutrino energy spectrum for {$\alpha_p^r=2.4$} in the upper panel of Figure \ref{fig:f5}, to guess $dL_{\nu}/dE_{\nu}$ for these nearby clusters. Then, the neutrino flux of each cluster can be calculated as
\begin{equation}
\frac{d\Phi_{\nu}}{dE_{\nu}} = \frac{1}{4\pi^2R_{\rm vir}^2}\frac{dL_{\nu}}{dE_{\nu}},
\end{equation}
where $R_{\rm vir}$ is the virial radius of the cluster.  Note that the above has the units of neutrinos ${\rm GeV^{-1}}$${\rm cm^{-2}}$${\rm s^{-1}}$ ${\rm sr^{-1}}$.

Figure \ref{fig:f7} shows $E_{\nu}^2d\Phi_{\nu}/dE_{\nu}$ as a function of $E_{\nu}$, predicted for the nearby clusters in Table \ref{tab:t1}, along with the IceCube flux \citep{aartsen2014} and the atmospheric muon and electron neutrino fluxes \citep[e.g.,][]{richard2016} for comparison. A few points are noticed. (1) Among the nearby clusters, the Coma, Perseus, and Ophiuchus clusters are expected to produce the largest fluxes. Yet, at $E_{\nu}=1$ PeV, the predicted fluxes are {$\lesssim10^{-4}$} times smaller than the IceCube flux. Hence, it is unlikely that high-energy neutrinos from clusters would be reckoned with IceCube, even after the stacking of a large number of clusters is applied. (2) At the neutrino energy range of several GeV to TeV, for which the flux data of the Super-Kamiokande detector are available \citep[see, e.g.,][]{hagiwara2019}, the fluxes from nearby clusters are smaller by {$\lesssim10^{-6}$} times than the atmospheric muon neutrino flux and smaller by {$\lesssim10^{-4}$} times than the atmospheric electron neutrino flux. Hence, it is unlikely that the signature of neutrinos from galaxy clusters could be separated in the data of ground detectors such as Super-Kamiokande and future Hyper-Kamiokande \citep[e.g.,][]{abe2011}. (3) Our neutrino fluxes from nearby clusters are substantially smaller than the ones estimated in previous works. For instance, our estimates are {$\sim10^{-3}$} times smaller than those for $\alpha_p=2.4$ at $E_{\nu}=250$ TeV in Table 3 of \citet{zandanel2015}. This discrepancy comes about mainly because our DSA model has a smaller acceleration efficiency, compared to the efficiency model adopted in their work (see Section \ref{sec:s2.3}), but also partly due to different approaches for modeling the CRp production in simulated clusters.

\section{Summary}
\label{sec:s4}

The ICM contains collisionless shocks of $M_s\lesssim5$, induced as a consequence of the LSS formation of the universe, and CRp are generated via DSA and then reaccelerated at the supercritical $Q_{\parallel}$ population of the shocks. Due to the long lifetime, the CRp are expected to be accumulated and mixed by turbulent flow motions in the ICM during the cosmic history. Then, inelastic CRp-p collisions should produce neutral and charged pions, which decay into $\gamma$-rays and neutrinos, respectively.

In this paper, we have examined the production of CRp in galaxy clusters and the feasibility of detecting $\gamma$-ray and neutrino emissions from galaxy clusters. To that end, we performed cosmological LSS structure simulations for a $\Lambda$CDM universe. In the post-processing step, we have identified shocks formed inside the virial radius of 58 simulated sample clusters, and measured the properties of shocks, such as the Mach number, the kinetic energy flux, and the shock obliquity angle. Adopting the model proposed in Paper I for fresh-injection DSA and a simplified model for reacceleration based on the test-particle solution, we have estimated the volume-integrated momentum distribution of CRp, produced by ICM shocks inside simulated clusters. Because we did not self-consistently follow the transport of CRp in simulations, we have assumed the radial distribution of the CRp density that scales with the gas density as $n_{\rm CR}(r,p)\propto{\bar n_{\rm gas}(r)}^{\delta}$ with $\delta = 0.5-1.0$. Then, we have calculated $\gamma$-ray and neutrino emissions from simulated clusters by adopting the approximate formalisms described in \citet{pfrommer2004} and \citet{kelner2006}, respectively. 

The main results of our study can be summarized as follows:\hfill\break
1) Inside simulated clusters, $\sim30$ \% of identified shocks are $Q_{\parallel}$, and $\sim23$ \% of the shock kinetic energy flux at $Q_{\parallel}$-shocks is dissipated by supercritical shocks with $M_s\ge2.25$. As a result, only $\sim7$ \% of the kinetic energy flux of the entire shock population is dissipated by the supercritical $Q_{\parallel}$-shocks that are expected to accelerate CRp. The fraction of the shock kinetic energy transferred to CRp via fresh-injection DSA is estimated to be {$\sim(1-2)\times10^{-4}$}.\hfill\break
2) The CRp, produced via fresh-injection DSA at supercritical $Q_{\parallel}$-shocks, have the momentum distribution, well fitted to a power-law. The volume-averaged power-law slope is $\alpha_p\sim2.4-2.6$, indicating that the average Mach number of CRp-producing shocks is $M_s\sim2.8-3.3$, which is typical for shocks in the cluster outskirts.\hfill\break
{3) Reacceleration due to the multiple shock passages of the ICM plasma makes the CRp spectrum harder. After the energization through reacceleration is incorporated in our model, the volume-averaged power-law slope reduces to $\alpha_p^r\sim2.35-2.5$, that is, the CRp spectrum flattens by $\sim0.05-0.1$ in slope. At the same time, the total amount of CRp energy contained in sample clusters increases by $\sim40-80$\%.}\hfill\break
4) The predicted $\gamma$-ray emissions from simulated clusters are mostly below the Fermi-LAT upper limits for observed clusters \citep[][]{ackermann2014}. Our estimates are lower than those of \citet{vazza2016} based on the DSA model of \citet{caprioli2014}, because our DSA efficiency, $\eta$, is smaller than their $\eta_{\rm CS14}$ in the range of $M_s=2.25-5$.\hfill\break
5) The predicted neutrino fluxes from nearby clusters are smaller by {$\lesssim10^{-4}$} times than the IceCube flux at $E_{\nu}=1$ PeV \citep{aartsen2014} and smaller by {$\lesssim10^{-6}$} times than the atmospheric neutrino flux in the range of $E_{\nu}\leq1$ TeV \citep{richard2016}. Hence, it is unlikely that they will be observed with ground facilities such as IceCube, Super-Kamiokande, and future Hyper-Kamiokande.

\acknowledgments
{We thank the anonymous referee for constructive comments that help us improve this paper from its initial form. We also thank Dr. K. Murase for comments on the manuscript.} J.-H.H. and D.R. were supported by the National Research Foundation of Korea (NRF) through grants 2016R1A5A1013277 and 2017R1A2A1A05071429. J.-H.H. was also supported by the Global Ph.D Fellowship of the NRF through 2017H1A2A1042370. H.K. was supported by the Basic Science Research Program of the NRF through grant 2017R1D1A1A09000567. A part of this work was done at KITP and supported by the National Science Foundation under grant NSF PHY-1748958.

\appendix

\section{Formulae for reacceleration in the Test-Particle Regime}
\label{sec:sa}
\restartappendixnumbering{}

{
If the preexisting CRp, upstream of shock, has a power-law spectrum, $f_{\rm pre}(p) = f_0\left({p}/{p_{\rm inj}}\right)^{-s}$,
the reaccelerated, downstream spectrum is given as
\begin{eqnarray}
f_{\rm reacc}^{(1)}(p) = \left\{\begin{array}{lr}
\left[{q/(q-s)}\right]\left[1-({p/p_{\rm inj}})^{-q+s}\right]f_{\rm pre}(p),
&\mbox{ if $q\neq s$},\\
q\ln(p/p_{\rm inj})f_{\rm pre}(p), &\mbox{ if $q= s$},
\end{array} \right. \nonumber\\ 
~~~
\label{a1}
\end{eqnarray}
where $q=3\chi/(\chi-1)$ with the shock compression ratio, $\chi$, is the test-particle power-law slope \citep{kang2011}. 

If the momentum spectrum of the preexisting CRp for the subsequent shock passage is taken as $f_{\rm reacc}^{(1)}$ in Equation (\ref{a1}), then, after the second reacceleration episode, the downstream spectrum has the following analytic form:
\begin{equation}
f_{\rm reacc}^{(2)}(p) = \left\{\begin{array}{lr}
\{q^2/(q-s)^2-[q^2/(q-s)\ln(p/p_{\rm inj}) \\
-q^2/(q-s)^2](p/p_{\rm inj})^{-q+s}\}f_{\rm pre}(p), & \text{if}~q \neq s, \\
q^2/2~[\ln(p/p_{\rm inj})]^2f_{\rm pre}(p), & \text{if}~q=s. 
\end{array}\right. \nonumber\\ 
~~~
\label{a2}
\end{equation}
}

\end{document}